\documentclass[doublecol]{epl2} 
\usepackage{subfigure}

\title{Snell's law for particles moving on piecewise homogeneous two dimensional surface with linear boundaries}
\shorttitle{Snell's law for particles moving on $2$-dimensional surfaces} 

\author{Pratik Mandrekar \and Toby Joseph\thanks{E-mail: \email{toby@bits-goa.ac.in}}}
\shortauthor{P. Mandrekar \etal}
\institute{                    
 Birla Institute of Science and Technology, K.K.Birala Goa Campus, Goa, India\\
 }

\abstract{
We study the properties of least time trajectories for particles moving on a two dimensional surface which consists 
of piecewise homogeneous regions. The particles are assumed to move with different constant speeds on different regions 
and on the boundary between regions. The speed of the particle is assumed to be highest when it moves along the edges
formed by the boundary of two regions. We get an analogous behavior to Snell's Law of light refraction, but in a more
generalized form. The model could be used for studying properties of animal and insect trails which tend to form predominantly 
along edges. The model predicts three types of behavior for the trajectories near a corner forming edge: fully edge following, 
partial edge following and complete avoidance of the edge, which are indeed observed in natural ant trails.
}

\begin{document}

\maketitle

\section{Introduction}
The mathematical problem of finding the shortest or least time path has a long history. Heron of Alexandria ($1^{\rm st}$ century AD)
was the first to realize that the law of reflection of light gives the shortest (or least time) path from a point to another on the same 
side of a dividing line which touches that line exactly once. In the seventeenth century Pierre de Fermat used the principle of least 
time to explain Snell's Law of refraction of light at the interface of two media in which the light travels with different speeds. 
The path that light ray takes in a medium with varying refractive index can be quite complex and is what is responsible for such
phenomena as mirage and looming.

Consider the refraction of light at the boundary of two media having different refractive indices. When one solves for the path by
looking for the least time trajectory, the assumption is that the light either travels in one medium or the other but not in the
interface of the two. But, it is possible to think of situations where the objects whose trajectories one is interested in might move
along the interface as well, with a speed different from its value in the two media. The question to ask then is: How do the trajectories 
get modified with this additional freedom that the particle has? If we allow for particle motion along the boundaries, the least time 
trajectories one gets are markedly different. The study of animal and insect trail patterns is one possible area where
some of these trajectories might be relevant.

Trail formation is a strategy that many organisms employ for efficient execution of a collective task 
to be done. Trail formation in ants and termites are the most ubiquitous examples . Ants use well organized trails for 
transportation of food to nest as well as for shifting of colonies \cite{holldobler1990ants}. Another example of path optimization
can be seen in the amoeba tube formation to the food source \cite{toshiyuki2007}. Trail formation is a fairly evolved collective
endeavor that helps organisms to make the right choice (the most rewarding food source, for example), in the most economic 
manner (say, using the least 'costly' route from food source to nest) most of the times. It is thus interesting to model the shapes 
these trails tend to take. 

Ant trails typically tend to form along the shortest route from the nest to the destination, where for example there could be a
food source\cite{beckers1992}. But a notable exception to this behavior happens in the presence of structured guidelines or
edges \cite{klotz2000}. The ants then seem to go out of the way to accommodate appreciable amount of edges in its trail even 
when this would mean a longer route as compared to the straight line path from nest to food. The common example of this 
behavior being the trails formed inside and outside the buildings along the wall edges, crevices, tile boundaries, table edges etc.
It has been shown that in some species of ants, edge is used as a visual guide\cite{pratt2001use}, whereas in other species they 
use it thigmotactically (by touch)\cite{dussutour2005amplification}. One can model the behavior of ant trails by assigning different
costs  for travel on the surfaces and along the edges and trying to find the least cost trajectory. This is equivalent to the problem of
finding the least time trajectories for particles moving with different velocities on different surfaces and along the edges, 
where speed  of the particle is proportional to the inverse of the cost.

\section{The model} 
We shall study the particle motion on two dimensional surface which  is piecewise homogeneous. That is, the surface is partitioned
into separate regions in each one of which the particles move with a different speed. Additionally, we assume that the particle moves 
with a different speed if the motion is along the edges formed at the boundaries of two regions. The aim is to find the least time 
trajectories connecting two given points on such a surface. The problem for a most general distribution of regions is well defined 
but only numerically tractable. We shall look at couple of special and interesting cases where the trajectories are analytically solvable.  

\section{Trajectories in the presence of a single straight boundary}
\begin{figure}[htbp]
\onefigure[width=8cm]{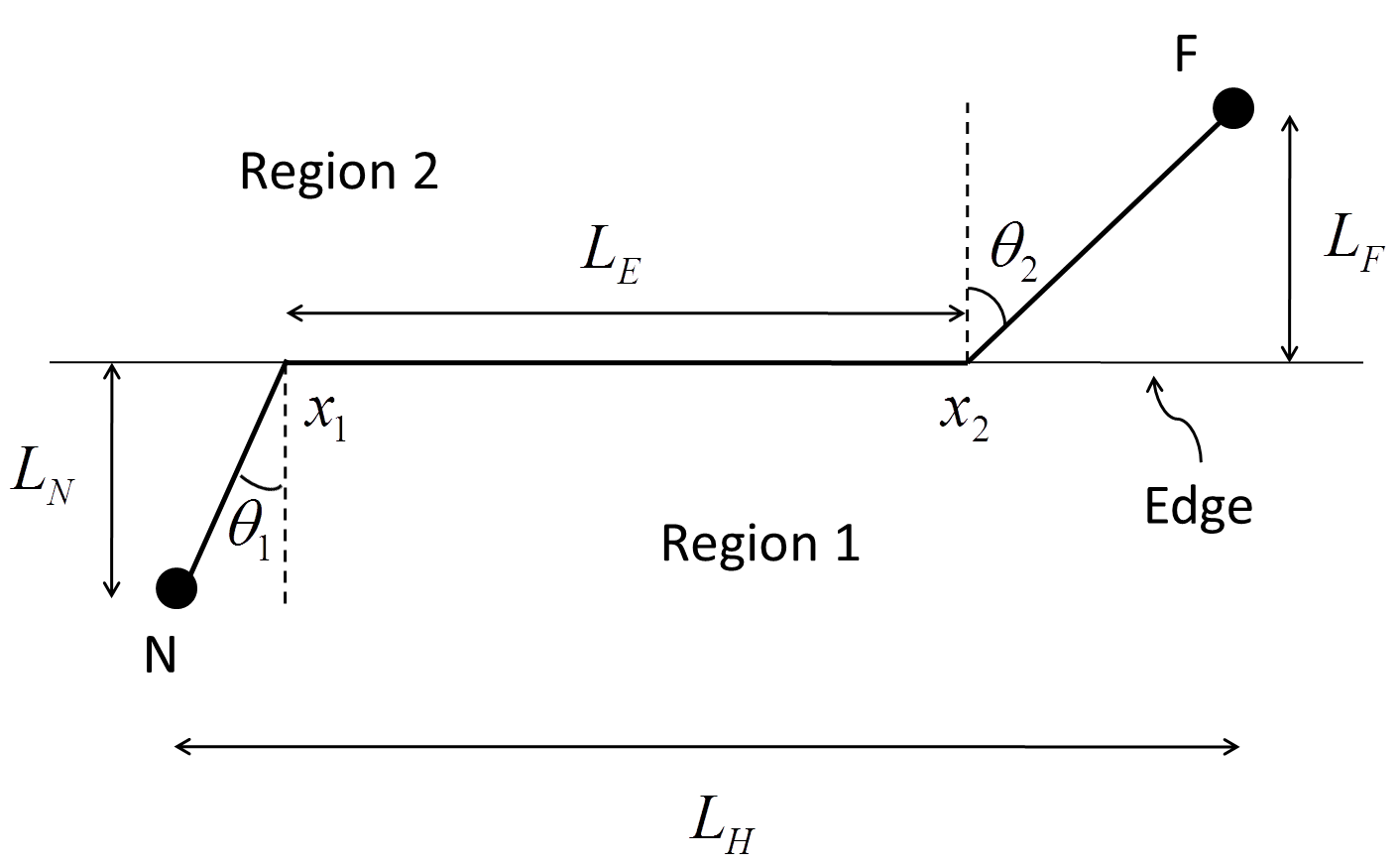}
\caption{Particle trajectory when the starting and the destination points are separated by a straight boundary.
Distances $x_1$ and $x_2$ are measured from the point $O$.}
\label{fig1}
\end{figure}

If there is only a single region present then the least time trajectory will be a straight line joining the starting to the
destination point. Let us consider a situation where two semi infinite regions ($S_1$ and  $S_2$) are separated by 
a straight boundary or edge (see fig. \ref{fig1}). The starting point($N$ in fig. \ref{fig1}) is in $S_1$ and the destination 
point is in $S_2$  ($F$ in fig. \ref{fig1}). Let $L_N$ and $L_F$ represent  the perpendicular distance from the edge to $N$ 
and $F$ respectively. The horizontal distance between $N$ and $F$ is represented by $L_H$. The particle velocities are 
$v_p$ when it moves along the boundary, and $v_1$ and $v_2$  when it moves on the regions $S_1$ and $S_2$ 
respectively. The quantities $C_i \equiv 1/v_i$  can be thought of as the cost of traversing unit distance on the two 
regions ($i = 1,2$) or the edge ($i = p$) and the least time problem can be equivalently posed now in terms of a least cost path. 
We shall take $C_p <  C_i \;(i = 1,2)$ as we assume that the particle has highest velocity when moving along edges.

Let $x_1$ and $x_2$ be the distances at which the particle enters and leave the edge respectively (see fig. \ref{fig1}), 
measured from the point $O$ in the figure. Let $\theta_1$ be the angle the trajectory makes with the perpendicular at $x_1$ 
and $\theta_2$ the angle the trajectory makes with the perpendicular at $x_2$.
The total cost associated with the trajectory, $C_T$, is
\begin{eqnarray}
C_T&=&\sqrt{(L_N \tan \theta_1)^2+L_N^2} \; C_1+(x_2-x_1) C_p \nonumber \\ 
&&\hspace{2.1cm}+\sqrt{(L_F\tan \theta_2)^2 + L_F^2}\; C_2
\label{tot_cost}
\end{eqnarray} 
Here we are assuming that $x_2$ is larger than $x_1$. Minimizing the cost $C_T$ with respect to the parameter $\theta_1$ 
and $\theta_2$ we get,
\begin{eqnarray}
\sin \theta_1 &=& \frac{C_p}{C_1}  \label{slaw1}\\
\sin \theta_2 &=& \frac{C_p}{C_2}
\label{slaw2}
\end{eqnarray}
This suggests that for fixed values of cost functions, entry angle in to the edge ($\theta_1$) and the angle at which the 
particle leaves the edge on to the second region ($\theta_2$) will be constants.  

The above analysis would break down if one ends up with a solution where $x_2< x_1$, since in that case the cost would be 
negative for the portion along the edge  (see eq. (\ref{tot_cost})). This would happen if 
\begin{equation}
\frac{L_N C_p}{\sqrt{C_1^2-C_p^2}} +\frac{L_F C_p}{\sqrt{C_2^2-C_p^2}} > L_H 
\end{equation} 
which can be obtained by noting that $x_1 = L_N \tan(\theta_1)$ and $x_2 = L_H - L_F \tan(\theta_2)$ and using the equations (\ref{slaw1}) and (\ref{slaw2}).
In such a case there exists no solution in which the trajectory has the edge as its part. The correct cost function to 
be minimized in such a case would be,
\begin{equation}
C_T=\sqrt{(L_N \tan \theta_1')^2+L_N^2} \; C_1 + 
\sqrt{(L_F\tan \theta_2')^2 + L_F^2}\; C_2  \nonumber
\end{equation} 
which is obtained by putting $x_2 = x_1$ in eq. (\ref{tot_cost}). The solution that minimizes this cost function is given by
\begin{equation}
\frac{\sin \theta_1'}{\sin \theta_2'}=\frac{C_2}{C_1}
\label{slawf}
\end{equation}
This gives the Snell's law for light refraction at the boundary of two surfaces. 

We see that the paths we have obtained is very similar to that of light ray refraction as the ray passes from one 
medium to another. The crucial difference in the present case being that the particle paths, unlike the light ray, can move 
along the interface of the  two media (edge). The angle $\theta_1'$ ($\theta_2'$) never exceeds $\theta_1$ ($\theta_2$), which 
is like the upper critical angle in the context of light refraction. But unlike in the case of light refraction where upper 
critical angle is present only for the ray that goes from a higher refractive index medium to a lower refractive index one, in the 
present case upper critical angle is always present, as  the speed travel along the edge is assumed larger than that for travel along
the surfaces.

\section{Trajectories in the presence of a corner forming boundary}
We now study the path that the particle takes when edges formed between the two regions is not linear.
\begin{figure}
\onefigure[width=8cm]{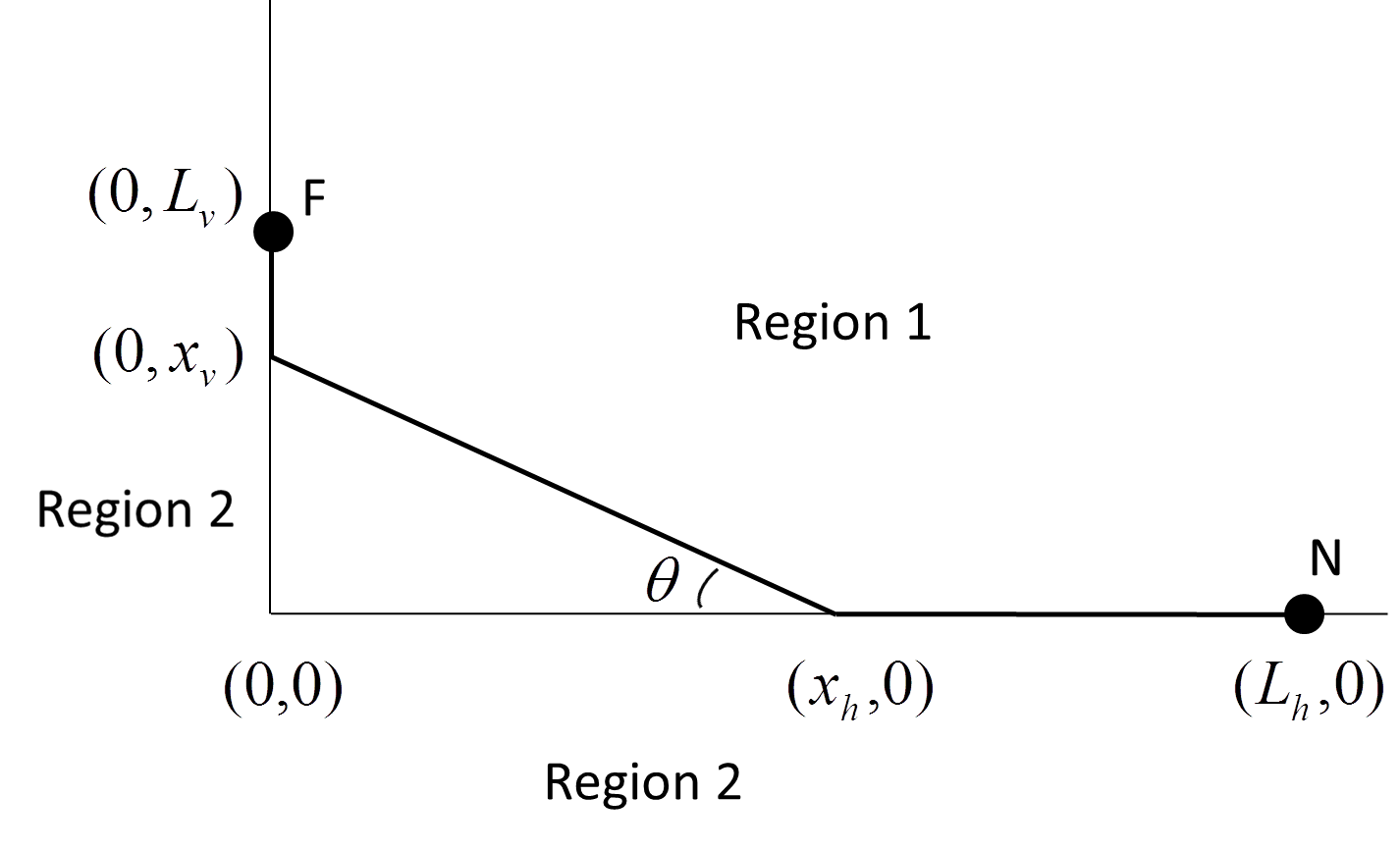}
\caption{Trajectories in the presence of corner forming boundaries. The region on the convex side (region 2)  has high cost of travel.
($C_2 \ge C_1$). The possible candidates for least energy trajectories are characterized by the two variables $x_v$ and $x_h$.}
\label{fig2}
\end{figure}
We shall consider a right angled boundary formed when the surface is divided into two parts with one quadrant being
formed by type $1$ region and the rest by type $2$ region as shown in fig. \ref{fig2}.  We find the least cost trajectory 
when the source and the destination points lie on the two perpendicular boundary edges. We shall again take 
$C_p$ to be less than $C_2$ and $C_1$. This problem has relevance to nature of ant trails formed on surfaces with corner 
forming guiding edges which we shall discuss in the next section.
 
Taking the corner point ($O$ in the fig. \ref{fig2}) as origin, $L_h$ and $L_v$ are the distances to the source and destination 
measured from the origin. Since $C_p \le C_2$ the minimum cost trajectory can have no part passing through region $2$. 
The various possible trajectories can thus be parametrized using two points, $x_h$ and $x_v$. These are the point 
of departure from edge on which source point lies on to region $1$ and point of entry into the perpendicular edge from region
$1$. The minimum cost trajectory should have $0 \le x_h \le L_h$ and $0 \le x_v \le L_v$ . If $C_p$ and $C_1$ represent 
the costs per unit length for traveling along the edges and along region $1$ respectively, the total cost incurred for a given 
path is given by, 
\begin{equation}
C_{b}=C_p(L_h-x_h)+C_p(L_v-x_v)+\sqrt{x_h^2+x_v^2}\;\; C_1
\label{cost_corn}
\end{equation}
where, $0 \le x_h \le  L_h$ and $0 \le x_v \le L_v$.

Minimizing the cost $C_{b}$ with respect to the parameter $x_h$ and $x_v$ one gets,
\begin{equation}
\frac{C_1}{C_p}= \frac{\sqrt{x_h^2+x_v^2}}{x_h} =  \frac{\sqrt{x_h^2+x_v^2}}{x_v} \;. \nonumber
\end{equation}
This tells us that if $X \equiv \frac{C_1}{C_p} = \sqrt{2}$, there is an extremum for all $x_h = x_v$ . The eigenvalues of the Hessian
matrix evaluated at the points $x_h = x_v$  are found to be $0$ and $\frac{C_1}{\sqrt{2} x_h}$. The eigenvector corresponding 
to the eigenvalue $0$  is in the direction given by $x_v = x_h$. This means that for $X = \sqrt{2}$,  there are a collection 
of paths all having the same costs and
with $x_v = x_h$. For other values of the ratio $\frac{C_p}{C_1}$ no solution exists in the interior region defined by 
$0 < x_h < L_h$ and $0 < x_v < L_v$. This implies that the minimum energy path in the $x_h - x_v$ plane should lie at the 
boundaries given by $x_h = L_h \; ,0 < x_v \le L_v$ or $x_v = L_v \; ,0 < x_h \le L_h$ or $x_h = x_v = 0$.
Note that if either $x_v$ or $x_h$ is zero, the trajectory corresponds to the one fully along the edge.

In order to find the minima when $X \ne \sqrt{2}$ we separately minimize the cost on the  boundary regions given above. 
The cost incurred when $x_v = L_v$ is given by,
\begin{equation}
C_b'=C_p(L_h-x_h)+\sqrt{x_h^2+L_v^2} \; C_1 \;. \nonumber
\end{equation}
Minimizing the above cost function with respect to $x_h$ we get,
\begin{equation}
x_h=\frac{L_v}{\sqrt{\frac {C_1^2}{C_p^2}-1}};\;\;
C_{min}'=C_p \left(L_h+L_v \sqrt{\frac {C_1^2}{C_p^2}-1}\; \right) \; \nonumber
\end{equation}
where $C_{min}'$ is the corresponding minimized energy. Similarly, substituting $x_h = L_h$ in Eq. (\ref{cost_corn}) and minimizing 
the cost function with respect to $x_v$ one gets the possible paths which minimizes the cost function given by,
\begin{equation}
x_v=\frac{L_h}{\sqrt{\frac {C_1^2}{C_p^2}-1}} ; \;\; 
C_{min}''=C_p \left(L_v+L_h \sqrt{\frac {C_1^2}{C_p^2}-1}\; \right) \nonumber
\end{equation}
where $C_{min}''$ is the corresponding minimized energy.
The cost incurred when the particle moves completely on the edge ($x_h = x_v = 0$) is $C_{min}^e = C_p(L_v + L_h)$.

The trajectory will deviate from the edge if one of the minimum costs computed above, $C_{min}'$ or $C_{min}''$, is less than 
the cost incurred while traversing the trail fully along the edge, $C_{min}^e$. The condition for the trail to deviate from the edge is
thus given by, $\frac{C_1}{C_p} < \sqrt{2}$. Further, if we assume that $L_h > L_v$ (the other case can be treated similarly), 
the solution for which $x_h = L_h$ is not valid because  $x_v = L_h/{\sqrt{C_1^2/C_p^2-1}}$  becomes larger than 
$L_v$. This definitely cannot be a minima. This spurious minima turns up because the cost function that we are using is not appropriate
when $x_v >  L_v$ or $x_h > L_h$. Thus for $L_h  > L_v $, the solution will have $x_v = L_v$ and $0 \le x_h \le L_h$.
For values of $X$  less than $X = \sqrt{1 + L_v^2/L_h^2}$ the value for $x_h$ that minimizes the energy will become larger 
than $L_h$ and thus unphysical. This implies that for those values of $C_1/C_p$, the minimum energy path would be given
 by $x_h = L_h$ and $x_v = L_v$.  

\begin{figure}
\onefigure[width=8cm]{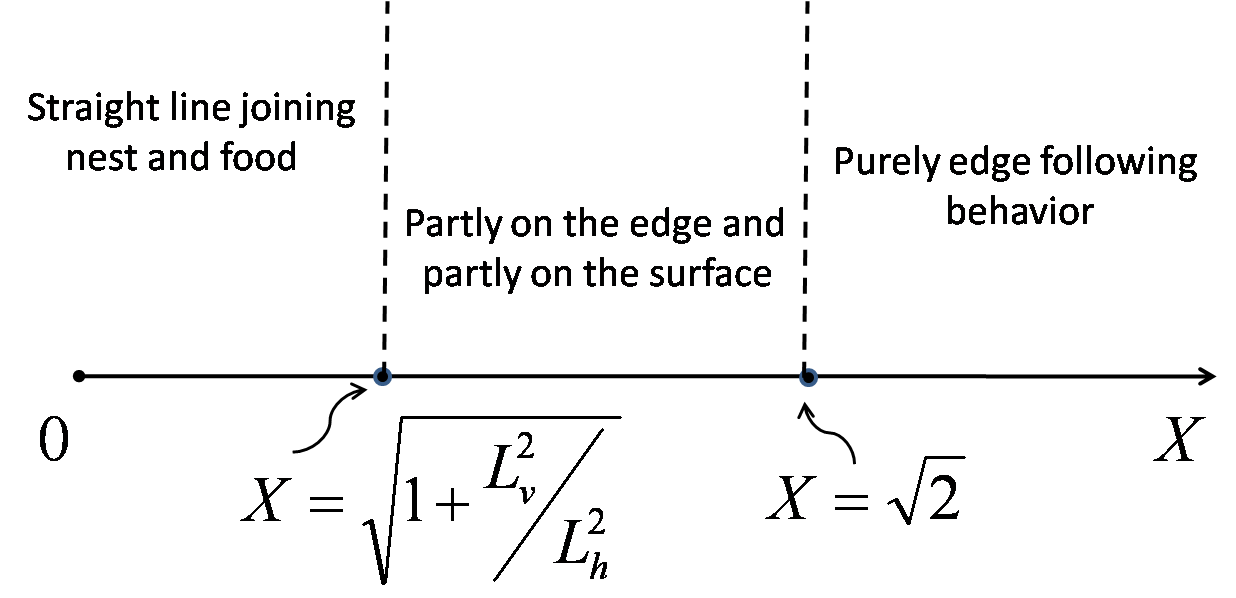}
\caption{The phase diagram showing various kinds of trajectories formed in the presence of corner forming edges assuming
$L_h > L_v$}
\label{fig3}
\end{figure}
The results we have derived above is shown in the phase diagram (see fig. \ref{fig3}). The phase diagram shown is for the case 
when $L_h \geq L_v$. For values  of $X$ above $\sqrt{2}$, the minimum energy path is the one where the trail is
along the edge completely. For values of $X$ lying between ${\sqrt 2}$ and $\sqrt{1 + L_v^2/L_h^2}$, the minimum energy 
path is partly along the horizontal edge and the rest on region $1$. For values of $X$ less than $\sqrt{1 + L_v^2/L_h^2}$, 
the most economic path is a straight line connecting source to destination. In the limit $L_h \gg L_v$, the range of $X$ 
values over which one can observe a path that is partially on the edge and partially on the surface is maximum. In the limit
of $L_h = L_v$, the path is either completely along the edge or completely on the surface depending on whether the value of 
$X$ is larger that $\sqrt{2}$ or not.

It is possible to generalize the above analysis to the case where the starting and the destination points do not lie on the edges
themselves. In this case the trajectories which has edges as its parts will then have to be characterized with more number of
parameters and one has to compare the cost of such paths with the cost of the straight line path between the two points.
We shall discuss below certain observations made on natural ant trails where the model and the particular case we have
looked at can qualitatively explain the features seen.

\section{Qualitative comparison with naturally observed trails}
\begin{figure*}
\begin{center}
\subfigure[]
{
\includegraphics[width=6cm]{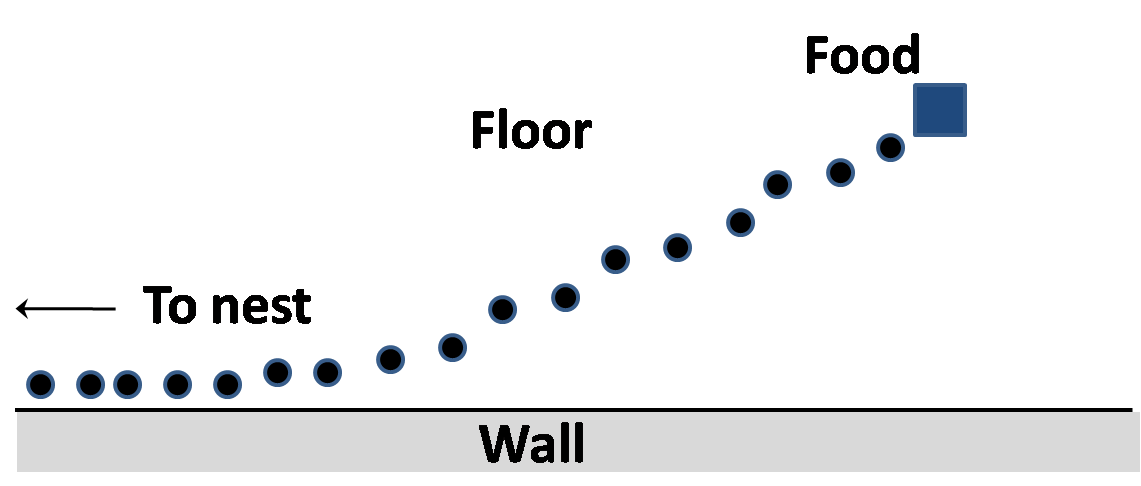} 
\label{fig4a}
}
\hspace*{1.5cm}
\subfigure[]
{
\includegraphics[width=6.2cm]{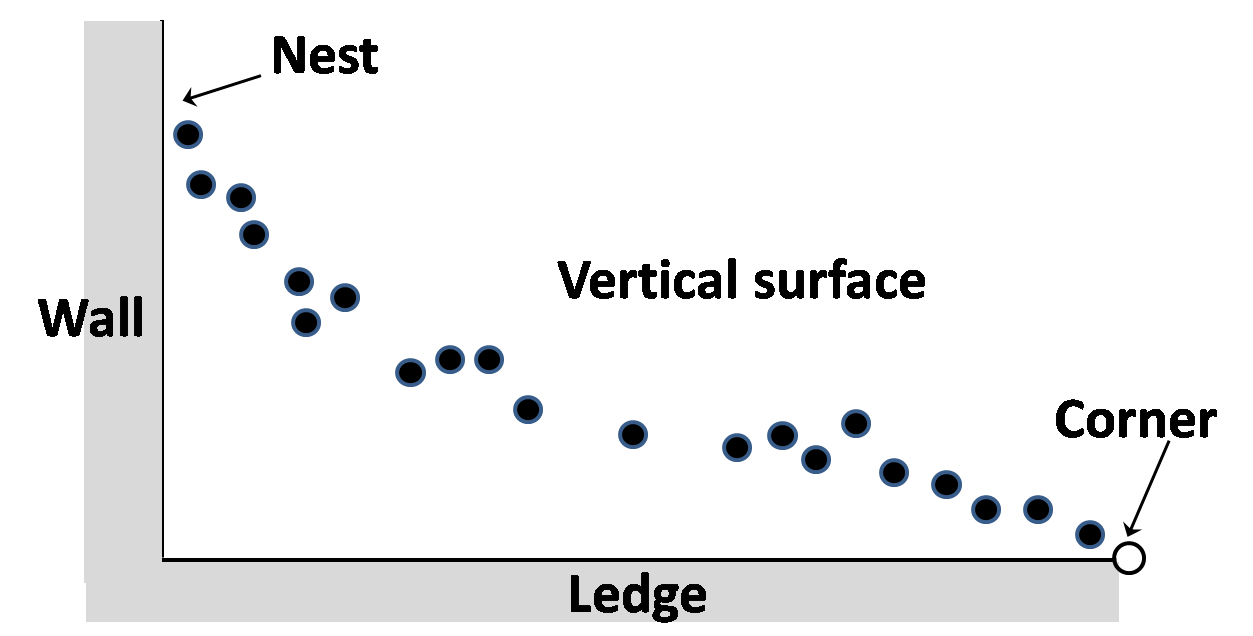}
\label{fig4b}
}
\hspace{0.5cm}
\subfigure[]
{
\includegraphics[width=6cm]{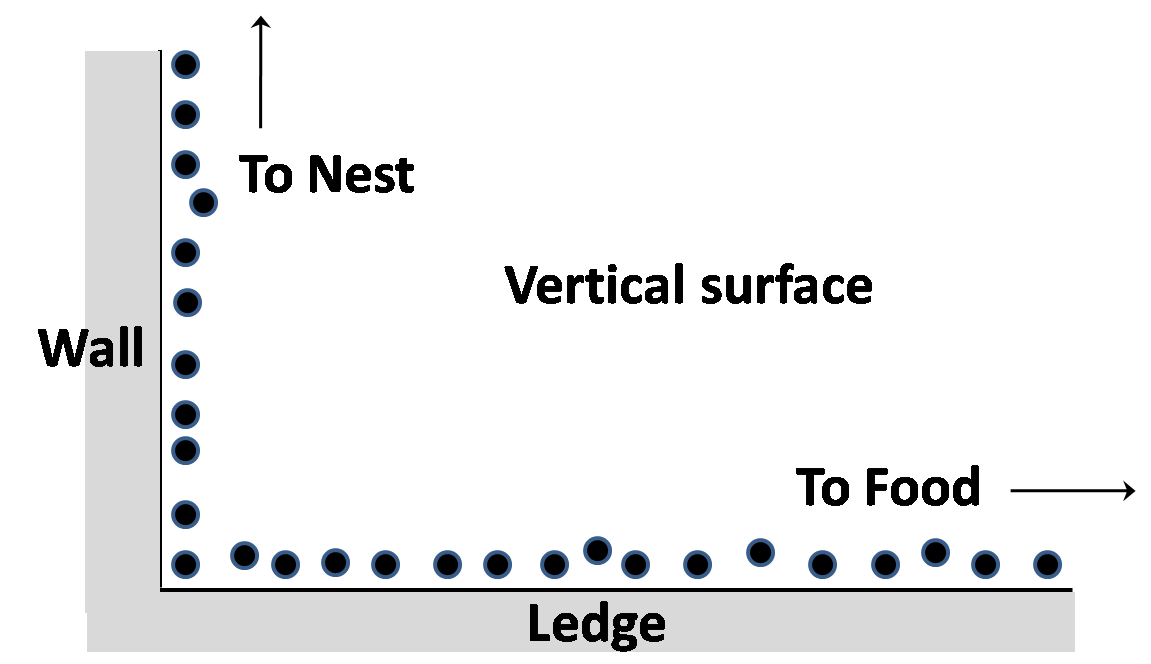}
\label{fig4c}
}
\hspace{1.5cm}
\subfigure[]
{
\includegraphics[width=6cm]{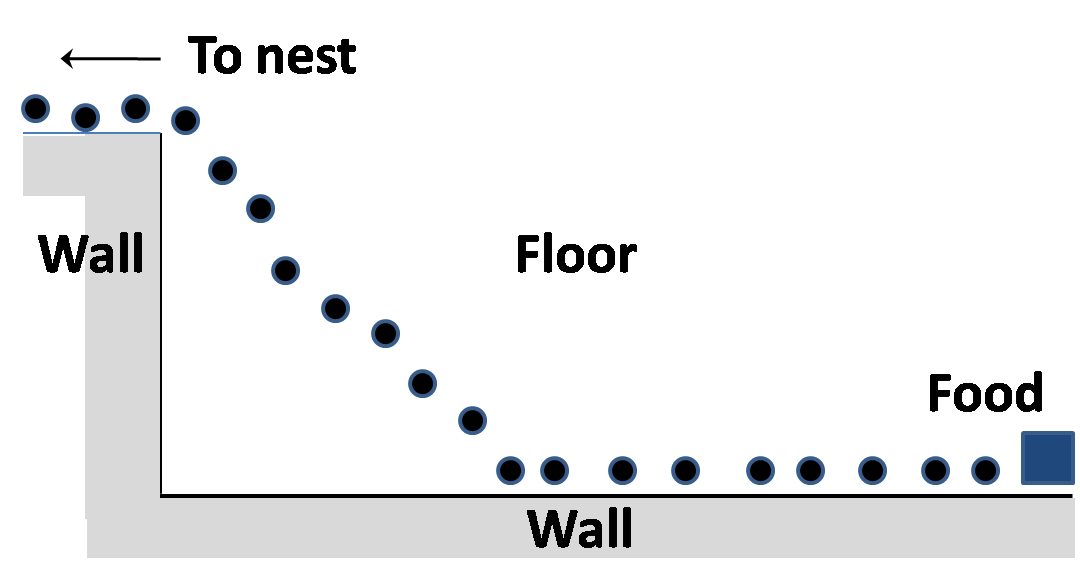}
\label{fig4d}
}
\label{fig4}
\caption{Different types of ant trails observed near wall edges and corners. The shaded regions in the figures correspond to surfaces coming out of the plane of the paper. (a) The trail formed along an edge cuts at an angle  ($\approx 24^o$) to the edge to 
access the food lying on the floor. (b) The ant trail completely avoids the edges and forms directly from the nest to the destination
($L_h = 13 {\rm cm}, L_v = 9 {\rm cm}$). (c) The case where the trail was seen to follow the edge completely. (d) Trail is formed
partially along the edge and partially on the surface ($L_v = 9 {\rm cm}$, $L_h = 50 {\rm cm}$). } 
\end{center}
\end{figure*}

 We have made some preliminary observations on the edge following feature of ant trails (see figs. \ref{fig4a},\ref{fig4b},\ref{fig4c}
and \ref{fig4d}) and they are qualitatively found to exhibit the types of behavior that has been seen in the analysis above. 
The ant trails have been photographed and then traced. Even though the locations of the ants come with a small error 
($\pm 1{\rm cm}$) due to visual copying of the ant locations from the photographs on to the figures, the overall feature of 
the trails are accurate. Some of the trails given in the figures were formed naturally,  but in couple of the cases (fig. \ref{fig4a} 
and fig. \ref{fig4d}), we introduced the food sources to let the trail form in the regions of interest. 

The most common situation that one encounters is when there is a food source on the floor, close to a straight edge at the intersection 
of a wall and the floor. The ant trails will typically take off at an angle from the edge and approach the food in a straight line as seen in 
fig. \ref{fig4a}. The situation here is much like the case that we have initially analyzed (see fig. \ref{fig1}) above with the starting
point lying on the boundary. The most economic path that the ants choose is not the shortest route but a route that partially uses the
edge (boundary) for navigation.  Let us now look at a few trail patters in the presence of corner forming non straight edges. 
In fig. \ref{fig4b} we have shown an ant trail that was seen to avoid the edges completely and form a straight line path from 
the nest to destination directly. 
The destination in the present case was the corner of the ledge, beyond which the trail 
was along the ledge in a direction perpendicular (into the paper in the figure) to the surface shown. Fig. \ref{fig4c} shows the case
where the trail stuck to the edge completely while navigating the corner. This is one of the phases (when $X > \sqrt{2}$) that is expected in the analysis. The trail shown in fig. \ref{fig4d} is typical of a situation where a long wall meets a short one leading to
two perpendicular edges meeting with $L_h \gg L_v$(in the present case  $L_v = 9 {\rm cm}$ and $L_h \approx 50 {\rm cm}$). 
The ant trail is seen to have formed partially on the edge and partially on the surface, conforming to the phase in fig. \ref{fig3} 
when $X < \sqrt{2}$ and $X > 1$. Occasionally, one also comes across trails where $x_v$  is not equal to $L_v$ or $0$ 
(with $L_h > L_v$), which contradicts our analysis. This could be due to the fact that the local rules that ants employ to find 
the minimum cost trail does not always lead to the path with least cost. 

Ideally, an independent calculation of $X$ should be done to make a comparison between the model and the experimental data. 
That would require experiments to be done with complete control over the type of cues that ants could use in the trail 
construction and maintenance. There are indications of such results in experiments that were done in the context of 
nocturnal orientation of black carpenter ants by Klotz et al.\cite{klotz1993nocturnal}. In the arena test conducted in this work, 
it was seen that shielding visual cues progressively lead to more and more edge following thigmotactic behavior. Experimental
verification of the phase diagram we are proposing can be done using similar experimental settings.
Another instance where a Snell's law like behavior has been seen is in the context of light avoiding amoeba path formation 
towards food source \cite{toshiyuki2007}. 

\section{Summary}
We have started with a very simple model, reminiscent of Hamilton's principle for
light rays, to study the nature of particle trajectories in a surface formed by different regions on which the particles move with different speeds. The particles are assumed to move along the edges formed by the intersecting surfaces with a different speed.
This leads to a generalized form of Snell's Law in the context of these particle paths. In the case where the boundaries are not
straight, the analysis shows that non trivial trajectories can be formed: from completely edge following to partial edge following
and complete avoidance of the edges. Our observations of real ant trails confirms the existence of these various phases. The
comparisons we have done are qualitative but quantitative study of the phase diagram we find is possible. 
In the context of ant trails, we have not discussed about the methods by which ants could be zeroing in on the optimal trail.
This has to be a set of local rules that ants employ to arrive at better routes \cite{bruckstein1993ant,couzin03a}. 
How these local rules are affected in the presence of linear or piece wise linear guidelines will be an interesting question to 
address. The ideas presented here can be of use in autonomous vehicle design which could use edges for navigation and for 
optimizing the functions of chemical trail laying ant robots \cite{wagner1999distributed}. That one could be seeing glimpses 
of something as fundamental as Snell's law, in fact a more generalized version of it, in something as ubiquitous as ant trails is 
both surprising and interesting.

\end{document}